\documentclass[11pt,
intlimits,sumlimits,
]{scrartcl}

\usepackage[utf8]{inputenc}
\usepackage{dsfont}
\usepackage{amsmath}
\usepackage{subfig}
\usepackage{graphicx}
\usepackage{color}
\usepackage[onehalfspacing]{setspace}

\usepackage[english]{babel}
\usepackage[babel]{csquotes}

\DeclareMathOperator{\diag}{diag}		
\DeclareMathOperator{\erf}{erf}			
\DeclareMathOperator{\trace}{tr}		
\DeclareMathOperator{\ii}{i}			
\newcommand{\e}[1]{\exp \left( #1 \right)}	

\newcommand{\quaddot}{\quad .}
\newcommand{\quadcomma}{\quad ,}

\begin{document}
\title{Credit risk: Taking fluctuating asset correlations into account}

\author{Thilo A. Schmitt\footnote{thilo.schmitt@uni-due.de}, Rudi Sch\"afer and Thomas Guhr\\{\normalsize Faculty of Physics, University of Duisburg-Essen, Lotharstr. 1, 47048 Duisburg, Germany}}

\date{\today}

\maketitle

\begin{abstract}
In structural credit risk models, default events and the ensuing losses are both derived from the asset values at maturity. Hence it is of utmost importance to choose a distribution for these asset values which is in accordance with empirical data. At the same time, it is desirable to still preserve some analytical tractability. We achieve both goals by putting forward an ensemble approach for the asset correlations. Consistently with the data, we view them as fluctuating quantities, for which we may choose the average correlation as homogeneous. Thereby we can reduce the number of parameters to two, the average correlation between assets and the strength of the fluctuations around this average value. Yet, the resulting asset value distribution describes the empirical data well. This allows us to derive the distribution of credit portfolio losses. 
With Monte-Carlo simulations for the Value at Risk and Expected Tail Loss we validate the assumptions of our approach and demonstrate the necessity of taking fluctuating correlations into account. 
 
\end{abstract}

\section{Introduction}

The subprime crisis of 2007-2009 showed that many small defaulting debtors can drag down the whole economy~\cite{Hull2009}. With the collapse of Lehman Brothers the devastating consequences of failed credit risk management became apparent to the public. The recessions following such crises further fuel a rise in the default probability~\cite{Crouhy2000}.
During the past years many studies have pointed out the importance of better credit risk estimation and proposed different approaches, see Refs. \cite{bielecki2004credit,bluhm2003introduction,Duffie1999,Ibragimov2007,lando2008credit,mcneil2005quantitative,Heitfield2006,Glasserman2006,Mainik2012} for an overview.

The crucial problem is to determine the loss distribution for a large portfolio of credit contracts. The danger lies in the heavy right tail of the loss distribution. Here, the very large losses occur, for example large single events such as Enron or many small events as during the subprime crisis. Reducing this tail would increase the stability of the financial system as a whole. It is often argued that diversification will lower the risk of a portfolio. This claim is highly problematic. In reality, correlations between the asset values are very important in assessing the risk if we consider a portfolio of credit contracts, eg, in the form of collateralized debt obligations (CDOs). It has been shown that in presence of even weak positive correlations diversification fails to reduce the portfolio risk, see~\cite{Schonbucher2001,Glasserman2004} for first passage models and for the Merton model~\cite{Schafer2007,Koivusalo2011,Schmitt2014}.

Here, we generalize the Merton model for credit portfolios to take fluctuating correlations between the asset values into account. The fluctuations are due to the non-stationarity inherent in financial markets~\cite{Schmitt2013}. Importantly, the covariance and correlation matrix of asset values changes in time~\cite{Zhang2011,Song2011,Munnix2012,Sandoval2012}.
We assume that the asset values are distributed according to a correlation averaged multivariate distribution, which we recently introduced~\cite{Chetalova2013}. The validity of this assumption is verified by an extensive empirical study of the asset returns. 
Starting from this distribution we derive analytical results for the probability distribution of portfolio losses in case of homogeneous average correlations between the assets. This ensemble approach leads to a drastic reduction of the parameter space. We are left with two parameters, the average correlation between asset values and the strength of the fluctuations. 
The special case of zero average correlation has been previously considered~\cite{Munnix2011a}.
In addition, we are able to derive a limiting distribution for a portfolio containing an infinite number of assets. This contributes a quantitative reasoning to the limits of diversification.

Furthermore, we provide Monte-Carlo simulations for the general case of empirical correlation matrices studying the Value at Risk (VaR) and Expected Tail Loss (ETL). First, we show that it is reasonable to use a homogeneous average correlation matrix if the heterogeneous volatilities and drifts are taken into account. Second, the importance of modeling the fluctuations of the correlations is underlined by the simulations.

The paper is structured as follows: Section \ref{ch:pavg} introduces a correlation averaged asset value distribution for the Merton model. In Section \ref{ch:data} we show that this distribution fits the empirical data well and we determine the parameter for the distribution from empirical data. We calculate the average loss distribution in Section \ref{ch:lossdist} and present the implications to risk management in Section \ref{ch:results}.

\section{An ensemble approach: average asset value distribution}
\label{ch:pavg}

\label{ch:model}

We use the Merton model~\cite{Merton1974} as a basis. We consider a portfolio of $K$ credit contracts  and assume that the obligors are publicly traded companies.
It was Merton's idea to use the stock price $S_k(t)$ of each obligor $k$ as a proxy for the asset value $V_k(t)$ at time $t$. At the maturity time $T$ the obligor has to pay back the face value $F_k$ of the credit contract. The equity of the company can then be viewed as a European call option on its asset value with a strike price equal to the face value $F_k$. The face value consists of the money given to the obligor, interest and a risk compensation. In the Merton model, a default occurs if the asset value at maturity, $V_k(T)$, is below the face value $F_k$. 
The ensuing normalized loss is then given as 
\begin{align}
L_k = \frac{ F_k - V_k(T) }{ F_k } \Theta ( F_k - V_k(T) )\quaddot
\end{align}
The Heaviside step function $\Theta ( F_k - V_k(T) )$
ensures that the loss $L_k$ is always equal to or greater than zero.
The total loss of the credit portfolio is given by the sum of the individual losses $L_k$, weighted by their fraction $f_k$ in the portfolio, 
\begin{align}
L = \sum_{k=1}^K f_k L_k \quad, \quad f_k = \frac{ F_k }{ \sum_{i=1}^K F_i } \quaddot
\label{eq:def}
\end{align}
Since the losses are a direct result from the asset values at maturity, the portfolio loss distribution can be calculated as a filter integral,
\begin{align}
	p(L) & = \int_{[0,\infty)^K} d[V] \ g (V|\Sigma)  \delta \left( L - \sum_{k=1}^K f_k L_k \right)
	\label{eq:loss3}
\end{align}
Here, $g(V|\Sigma)$ is the distribution of the asset values at maturity time $T$ with $V=(V_1(T),\dots, V_K(T))$ and $\Sigma$ is a stationary $K \times K$ covariance matrix without fluctuations.
The measure $d[ V ]$ is the product of the differentials for each asset value $V_k(T)$.
Equation~(\ref{eq:loss3})  was cast into the form
\begin{align}
 p(L) = \frac{ 1 }{ 2 \pi } \int_{-\infty}^{+\infty} d\nu \ e^{-\ii \nu L} \prod_{k=1}^K \left( \int_0^{F_k} dV_k(T) \exp \left ( \ii \nu f_k \left( 1 - \frac{ V_k(T) }{ F_k } \right) \right) + \int_{F_k}^\infty d V_k(T) \right) g (V|\Sigma) \quaddot
\label{eq:loss}
\end{align}
in~\cite{Munnix2011a}.   

We have two requirements for the distribution of the asset values $g(V|\Sigma)$. First, it has to be realistic, ie, the distribution should describe the empirical data well and address the non-stationarity of financial markets. Second, it should be analytically tractable.

We achieve both goals by using an ensemble approach which allows us to greatly simplify the description of the financial market. In~\cite{Chetalova2013} we introduced the random matrix average of a correlated multivariate normal distribution, which takes the non-stationarity of the covariance matrix into account. Importantly, our ensemble is not fictitious, it really exists as a consequence of the non-stationarity: The correlation or covariance matrices measured at different times differ from each other at a statistically significant level. The set of all these matrices forms our ensemble. 
The distribution is the result of averaging a multivariate normal distribution over an ensemble of Wishart distributed correlation matrices~\cite{Wishart1928}. We use it to describe the return vector $r(t)=(r_1(t),\dots,r_K(t))$ with
\begin{align}
	r_k(t) = \frac{ S_k(t + \Delta t) - S_k(t) }{ S_k(t) } \quaddot
\label{eq:return}
\end{align}
of $K$ stocks, where $\Delta t$ is the return interval. We start with the average return distribution $\langle g \rangle(r | \Sigma, N)$ which is compared to the data comparison in section~\ref{ch:data}. To simplify the notation, we omit the time dependence of $r$ when $r$ appears in the argument of a distribution. The average asset value distribution can then easily be calculated from the average return distribution. 
The general result for the average return distribution with fluctuating covariances
is
\begin{align}
\langle  g \rangle(r|\Sigma,N) & = \frac{\sqrt{N}^K}
                                                                    {\sqrt{2}^{N-2}\Gamma(N/2)\sqrt{\det(2\pi\Sigma)}} \frac{\mathcal{K}_{(K-N)/2} \left(\sqrt{Nr^\dagger\Sigma^{-1}r}\right)}
            {\sqrt{Nr^\dagger\Sigma^{-1}r}^{(K-N)/2}}
\label{eq:genresult}
\end{align}
with the Bessel function $\mathcal{K}$ of second kind and of order $(K-N)/2$. We do not use the Wishart distribution to describe the measurement noise due to finite time series, but to model the actual non-stationarity of the covariances. Hence, $N$ is a free parameter, which controls the strength of the fluctuations around the average covariance matrix $\Sigma$, as discussed in more detail in Section~\ref{ch:data}. 
In the context of credit risk, we are interested in changes of the asset values over the time period $T$. Therefore we will consider an equally long return interval, 
\begin{align}
	\Delta t = T
\end{align}
In the following we express the covariance matrix $\Sigma$ in terms of the average correlation matrix $C$ with $\Sigma=\sigma C \sigma$, where $\sigma=\text{diag}(\sigma_1,\dots, \sigma_K)$ contains the standard deviations for each of the $K$ assets.

We achieve analytical tractability by assuming a homogeneous average correlation matrix, 
\begin{align}
C=(1-c) \mathds{1}_K + c ee^\dagger \quadcomma
\label{eq:C}
\end{align}
where $\mathds{1}_K$ is the $K\times K$ unit matrix and $e$ is a $K$ component vector with ones as its elements. This implies that all off-diagonal matrix elements are equal to $c$. We emphasize that this is only an assumption about the average correlation matrix, the correlations in our approach fluctuate around this mean value. Later on, we show that this approximation yields a good description of the empirical data. 
This construction allows us to capture the correlation structure of the financial market by only two parameters: the average correlation level $c$ and the parameter $N$ which indicates the strength of the fluctuations around this average.
Inserting the homogeneous correlation matrix (\ref{eq:C}) into the general result for the average return distribution (\ref{eq:genresult}) and performing the calculations described in~\cite{Schmitt2014} yields
\begin{align}
\langle g \rangle (V | c, N) & = \frac{ 1 }{ 2^{N/2} \Gamma(N/2) \det \rho } \left( \prod_{k=1}^K \frac { 1 }{ V_k(T)}  \right) \int_0^\infty dz \ z^{ N / 2  - 1 } e^{- z/2} \notag \\
& \qquad \times \sqrt{ \frac{ N }{ 2 \pi z } } \sqrt{ \frac{ N }{ 2 \pi z ( 1 - c) T } }^K  \int_{-\infty}^{+\infty} du \ \e{- \frac{ N }{ 2 z } u^2 }\notag  \\
& \qquad \times \exp \left( - \sum_{k=1}^K \frac { N \left( \ln \frac{ V_k(T) }{ V_{k0} } - ( \mu_k - \frac{ \rho_k^2 }{ 2 } )T + \sqrt{ c T } u \rho_k \right)^2 }{ 2 z (1-c)T \rho_k^2 } \right) \quaddot
\label{eq:avg_result}
\end{align}
with $V_{k0} = V_k(0)$. The term $\prod_{k=1}^K V_k^{-1}(T)$ is the result of the Jacobian determinant due to the transformation from returns $r_k$ to asset values $V_k(T)$
\begin{align}
r_k \longrightarrow \ln\frac{V_k(T)}{V_{k0}} - \left(\mu_k-\frac{\rho_k^2}{2}\right)T \quadcomma
\label{eq:trafo}
\end{align}
using It\=o's Lemma~\cite{Ito1944}. In the spirit of the Merton model we assume that the asset values follow a geometric Brownian motion. We recall that the maturity time sets the return interval, $\Delta t = T$. When we later insert the asset value distribution~(\ref{eq:avg_result}) into Equation~(\ref{eq:loss}), we want to achieve a factorization of the $V_k$ integrals. Therefore, we do not perform the $u$ integration at this point. Equation~(\ref{eq:avg_result}) is our result for the average asset value distribution at maturity time $T$.
Due to the coordinate transformation from returns to asset values, the formerly dimensionless standard deviations $\sigma_k$ are substituted by the volatilities $\rho_k$ with dimension one over the square root of time, 
\begin{align}
\sigma_k = \rho_k \sqrt{T} \quaddot
\end{align}
Before we derive the average loss distribution, we give an extensive empirical validation of our ensemble approach in the next section.

\section{Empirical return distribution}
\label{ch:data}

We carry out the data analysis on two different datasets, which are obtained from Yahoo Finance~\cite{yahoo}. We select stocks from the Standard \& Poor's 500--index (S\&P 500) which are continuously traded in each time period we study. The composition of the S\&P 500 is focused on the  performing US companies. The second dataset we use includes all stocks which are continuously traded at the NASDAQ during our observation periods. The NASDAQ dataset takes a much broader selection of companies into account.

\begin{figure*}[htbp]
  \begin{center}
    \includegraphics[width=0.85\textwidth]{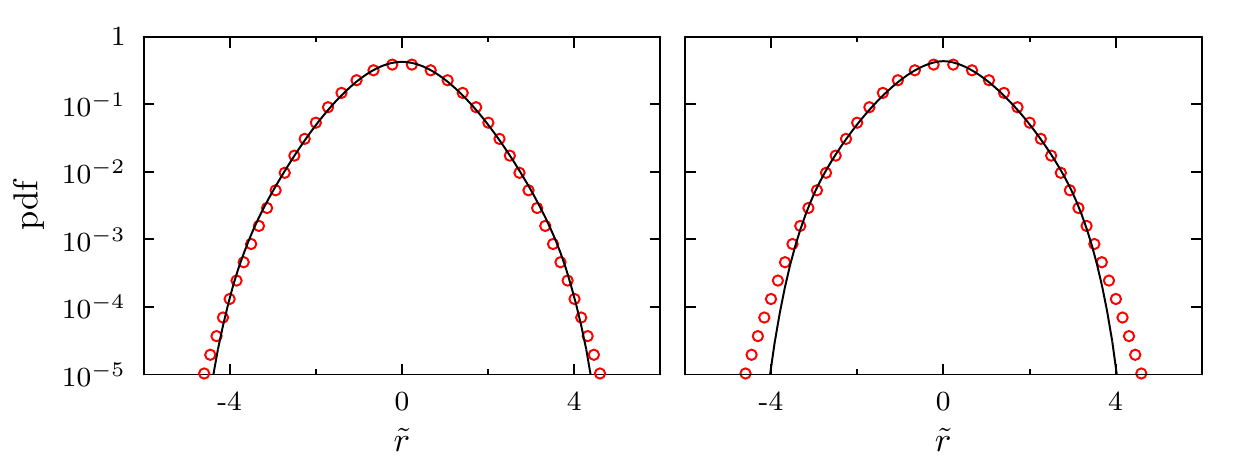}
  \end{center}
 \caption{Aggregated return distributions for fixed covariance matrices. The circles show a normal distribution. Left: the S\&P 500 dataset between 1992 to 2012 ($307$ stocks), right: the NASDAQ dataset from 2002 to 2012 ($2667$ stocks).}
 \label{fig:pw}
\end{figure*}

We analyse the multivariate time series of returns (\ref{eq:return}) for these two datasets.
For the derivation of the average return distribution $\langle  g \rangle(r|\Sigma,N)$, see Eq.~(\ref{eq:genresult}), we assume that the returns are well described by a multivariate normal distribution
\begin{align}
	g(r|\Sigma_\text{st})  = \frac{1}{\sqrt{\det(2\pi\Sigma_\text{st})}} 
    \exp\left( -\frac{1}{2} r^\dagger \Sigma_\text{st}^{-1}r\right)  \quadcomma
\label{multivar}
\end{align}
on short time horizons 
where the covariance matrix $\Sigma_\text{st}$ for this time interval can be viewed as stationary, see~\cite{Chetalova2013}. We recall that we receive the average return distribution by averaging Eq.~(\ref{multivar}) over an ensemble of Wishart distributed correlation matrices. Again, we notice that the correlation matrix is closely connected to the covariance matrix via $\Sigma = \sigma C \sigma$. 
We calculate the return time series for a return interval of $\Delta t=1$ trading day. 
These time series are then partitioned into short, non-overlapping intervals of 25 trading days.
On these short intervals we can view the covariance matrix $\Sigma_\text{st}$ as constant. Since the length of the time series on these intervals is smaller than the dimension of the covariance matrix, it is non--invertible. This will not cause problems from a mathematical point of view, because the distribution in Eq.~(\ref{multivar}) is properly defined in terms of $\delta$-functions.
For the data analysis we take all pairs of returns $(r_k,r_l)$ which, if our assumption~(\ref{multivar}) is true, should be bivariate normal distributed with a $2\times 2$ covariance matrix $\Sigma^{(k,l)}_\text{st}$. This matrix is always invertible. Next, we rotate the vectors $(r_k, r_l)$ into the eigenbasis of $\Sigma^{(k,l)}_\text{st}$ and normalize the elements of the vector with the corresponding eigenvalues. The aggregated distribution of all pairs of returns is shown in Fig.~\ref{fig:pw}. The left figure shows the aggregated results from a  dataset with $307$ stocks taken from the S\&P 500 index which where continuously traded in the period from 1992-2012. The right figure presents the distribution of all pairs of returns for a dataset consisting of $2667$ stocks from NASDAQ from 2002 to 2012. Both distributions are for daily returns. We find that for short time horizons, assuming a fixed covariance matrix $\Sigma_\text{st}$, the distribution of all pairs of returns fits a normal distribution rather well.

We now use the average return distribution (\ref{eq:genresult}) to calculate a distribution of rotated and scaled returns which can be compared to empirical data. We rotate the vector $r$ into the eigenbasis of the covariance matrix $\Sigma$ and normalize each element with its corresponding eigenvalue. Integrating out all degrees of freedom except one, we obtain the distribution of the rotated and scaled returns $\tilde{r}$
\begin{align}
\langle g \rangle(\tilde{r}|N) & = \frac { \sqrt{ 2 }^{1-N} \sqrt{ N } }{ \sqrt{\pi} \Gamma(N/2) } \sqrt{N \tilde{r}^2}^{(N-1)/2} \mathcal{K}_{(N-1)/2} \left(\sqrt{N \tilde{r}^2} \right)
\label{rk}
\end{align}
which once again contains a Bessel function of the second kind but now of order $(N-1)/2$.

We evaluate the probability distribution of the rotated and scaled returns from empirical data for the averaged covariance matrix $\Sigma$ and the covariance matrix $\hat\Sigma=\sigma C \sigma$ with homogeneous correlation structure, given the averaged correlation matrix $C$ as in Eq.~(\ref{eq:C}).

\begin{figure*}[htbp]
  \begin{center}
    \includegraphics[width=0.85\textwidth]{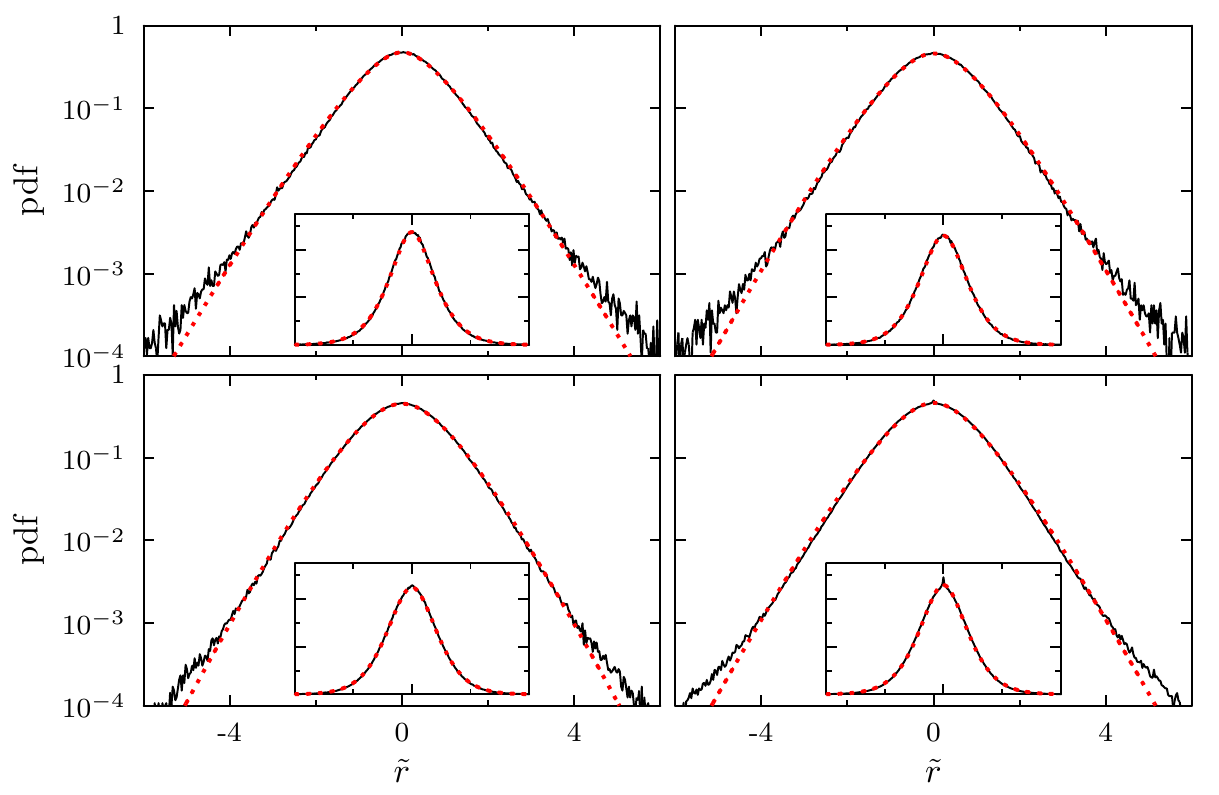}
  \end{center}
 \caption{Aggregated distributions for the rotated and scaled daily returns. The empirical distribution (black line) is compared to the theoretical result (red, dotted line). Top left/right: S\&P 500 (1992-2012) / (2002-2012), bottom left/right: NASDAQ (1992-2012) / (2002-2012). The insets show the corresponding linear plots.}
 \label{fig:cov1}
\end{figure*}

\begin{figure*}[htbp]
  \begin{center}
    \includegraphics[width=0.85\textwidth]{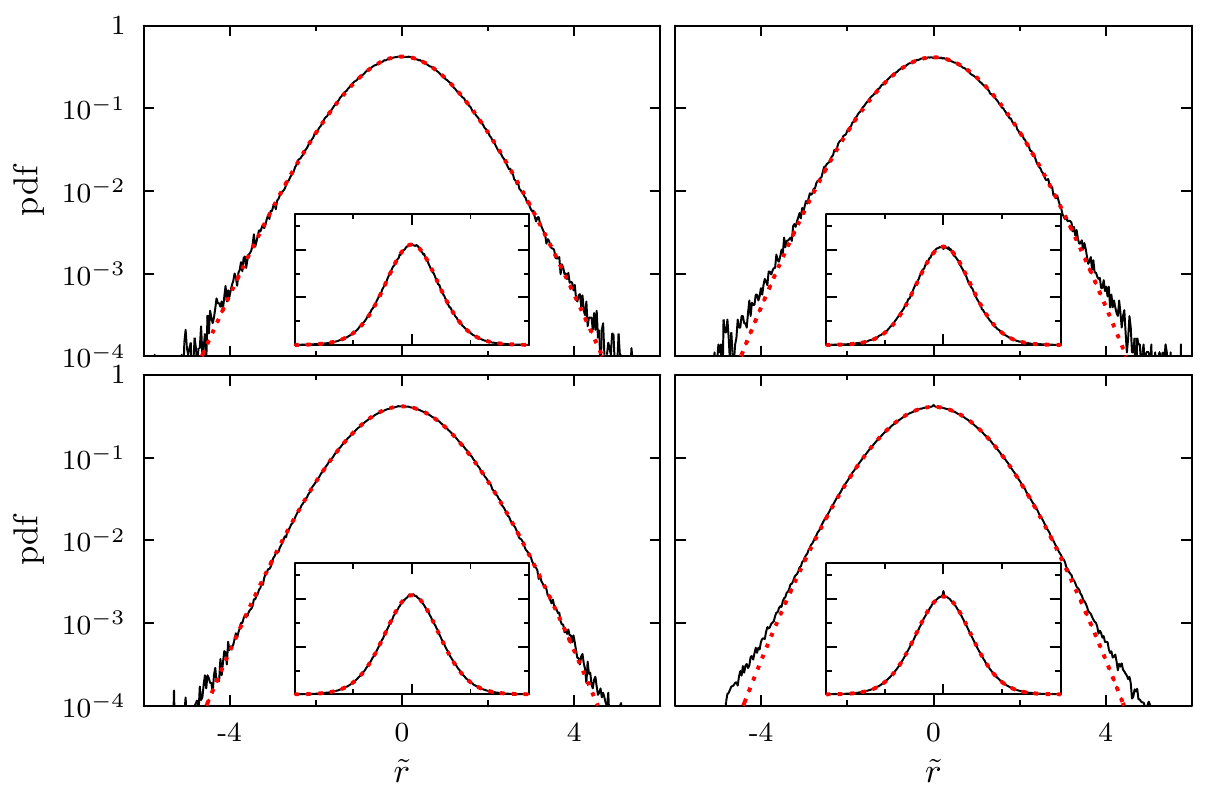}
  \end{center}
 \caption{Aggregated distributions for the rotated and scaled monthly returns. The empirical distribution (black line) is compared to the theoretical result (red, dotted line). Top left/right: S\&P 500 (1992-2012) / (2002-2012), bottom left/right: NASDAQ (1992-2012) / (2002-2012). The insets show the corresponding linear plots.}
 \label{fig:cov20}
\end{figure*}

Figures \ref{fig:cov1} and \ref{fig:cov20} show the probability distributions for the rotated and scaled returns calculated using the empirical covariance matrix $\Sigma$ with inhomogeneous correlation structure. We analyzed four different datasets on two time horizons. The first dataset consists of $307$ stocks taken from the S\&P 500 index which were continuously traded in the period from 1992-2012 (top left). The second set includes $439$ stocks from the S\&P 500 on the time horizon from 2002 to 2012 (top right). Then we use $708$ and $2667$ stocks from NASDAQ for the two time horizons (bottom left/right),  respectively. The results are shown in Fig.~\ref{fig:cov1} for daily returns and in Fig.~\ref{fig:cov20} for monthly returns. We carry out a Cramer-von Mises test to confirm the result of the least squares fit which determines the parameter $N$ in Eq.~(\ref{rk}), see Table~\ref{tab:n}.
 \begin{table*}[htbp]
\centering
\begin{tabular}{rrrrr}
\hline
& \multicolumn{2}{c}{S\&P 500} & \multicolumn{2}{c}{NASDAQ}\\
figure & 1992-2012 & 2002-2012 & 1992-2012 & 2002-2012\\
\hline
\ref{fig:cov1} & 5 & 6 & 7 & 6\\
\ref{fig:cov20} & 14 & 22 & 20 & 20\\
\ref{fig:mcov20} & 4 & 4 & 4 & 4\\
\hline
\end{tabular}
\caption{Values for the parameter $N$ used in Figs.~\ref{fig:cov1} to~\ref{fig:mcov20}.}
\label{tab:n}
\end{table*}
For the daily returns we find that values around $N=6$ describe the data well, while for monthly returns larger values around $N=20$ are necessary. In all cases  Eq.~(\ref{rk}) describes the data well.

\begin{figure*}[htbp]
  \begin{center}
    \includegraphics[width=0.85\textwidth]{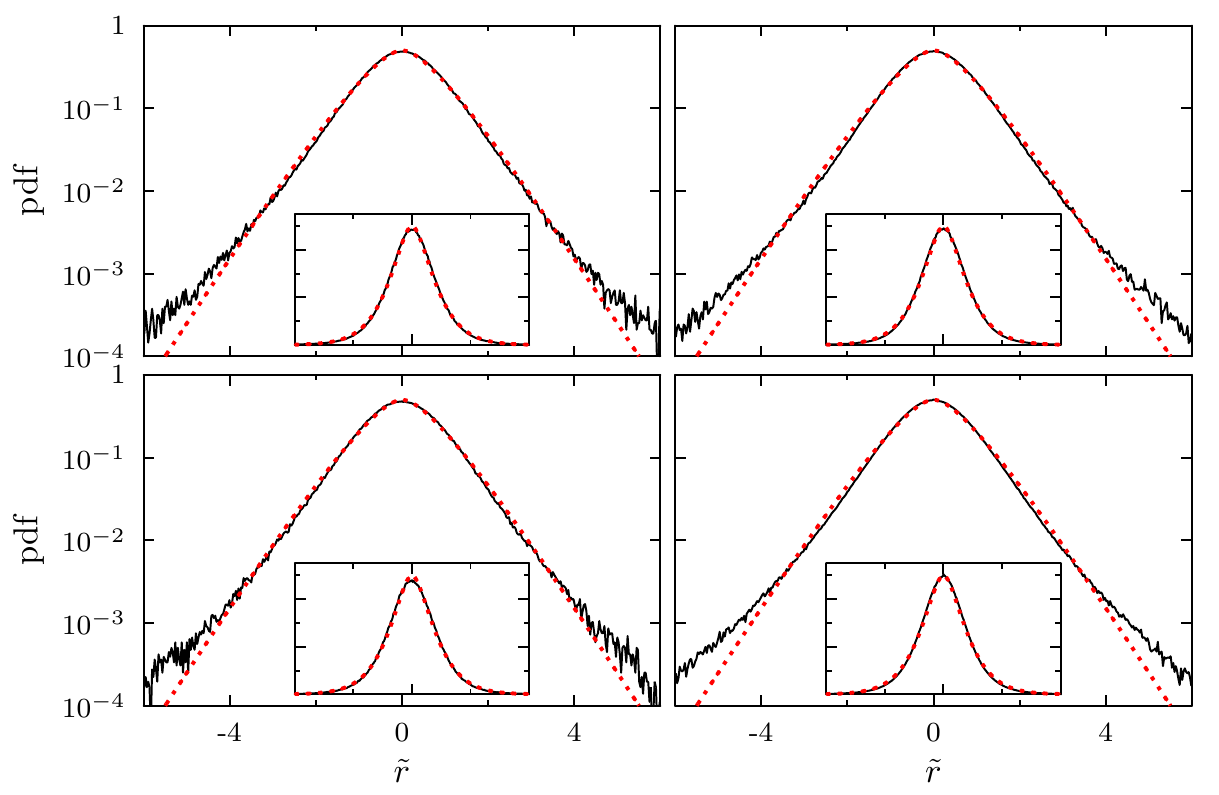}
  \end{center}
 \caption{Aggregated distributions for the rotated and scaled monthly returns using the covariance matrix with homogeneous correlation structure. The empirical distribution (black line) is compared to the theoretical result (red, dotted line). Top left/right: S\&P 500 (1992-2012) / (2002-2012), bottom left/right: NASDAQ (1992-2012) / (2002-2012). The insets show the corresponding linear plots. 
 The average correlation levels are $c=0.26$, 0.35, 0.21 and $0.25$, respectively.}
 \label{fig:mcov20}
\end{figure*}

Our approximation~(\ref{eq:C}) has an impact on the resulting values of $N$. To study it, we repeat the procedure discussed above but instead of using the empirical covariance matrix $\Sigma$ we evaluate the average correlation for the selected time horizon and substitute all off-diagonal  elements in the correlation matrix by this average. For the homogeneous matrix we find smaller values for $N$. In case of the S\&P 500 and NASDAQ $N=4$ is a good fit. The resulting probability distributions are shown in Fig.~\ref{fig:mcov20}. The top figures show the S\&P 500 data for the years 1992 to 2012 and from 2002 to 2012, the bottom figures show the NASDAQ data. We observe that smaller values around $N=4$ are needed to describe the distribution of monthly rotated and scaled returns if we use the covariance matrix with homogeneous correlation structure. This has to be compared to larger values around $N=20$ that we find for the covariance matrix with inhomogeneous correlation structure in Fig.~\ref{fig:cov20}. These results underline the interpretation of the parameter $N$ as an inverse variance. If our average correlation matrix has the same entries in all off-diagonal elements, a smaller value for $N$ is needed to describe the larger fluctuations around this average.

\begin{figure*}[htbp]
  \begin{center}
    \includegraphics[width=0.9\textwidth]{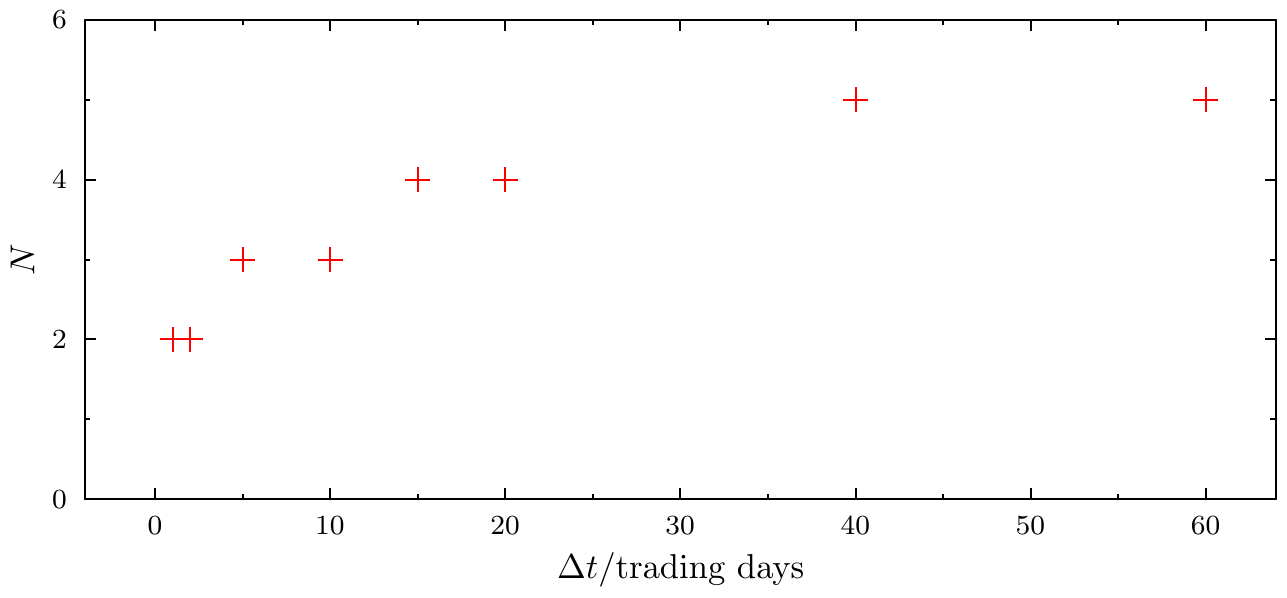}
  \end{center}
 \caption{Parameter $N$ versus the return interval $\Delta t$ in the case of a covariance matrix with homogeneous correlation structure.}
 \label{fig:mises_mcov}
\end{figure*}

Figure \ref{fig:mises_mcov} shows the $N$ values with the highest confidence using the Cramer-von Mises test for different return intervals $\Delta t$. For larger return intervals, $N=5$ gives the best fit to the data. 
An additional approach to determining the value of $N$ is discussed in appendix~\ref{ch:appendix:var}. It considers the variance of the expression $r^\dagger r$ and strongly corroborates the results of this section.

\section{Average loss distribution with fluctuating asset correlations}
\label{ch:lossdist}

Having justified our assumptions for the asset value distribution with empirical data, we now present the result for the average loss distribution which takes fluctuating asset correlations into account.

\subsection{General result for large portfolios}

Inserting the average asset value distribution (\ref{eq:avg_result}) into the loss distribution (\ref{eq:loss}) and performing an expansion for large $K$ as shown in~\cite{Schmitt2014},
we arrive at
\begin{align}
\langle p \rangle (L|c,N) & = \frac{ 1 }{ \sqrt{ 2 \pi } 2^{N/2} \Gamma( N / 2 ) } \int_0^\infty dz \ z^{N/2 -1} \text{e}^{-z/2}   \sqrt{ \frac{ N }{ 2 \pi } }\notag \\
& \qquad \times \int_{-\infty}^{+\infty} du \e{-\frac{ N }{ 2 } u^2 } \frac{ 1 }{ \sqrt{ M_2(z,u) } } \e{ - \frac{ ( L - M_1(z,u) )^2 }{ 2 M_2(z,u) } }
\end{align}
for the average loss distribution with
\begin{align}
M_1(z,u) =  \sum_{k=1}^K f_k m_{1k}(z,u)
\label{eq:M1}
\end{align}
and
\begin{align}
 \quad M_2(z,u) = \sum_{k=1}^K f_k^2 \left( m_{2k}(z,u) - m_{1k}^2(z,u) \right) \quaddot
 \label{eq:M2}
\end{align}
The $j$-th moments $m_{jk}(z,u)$ are
\begin{align}
m_{jk}(z,u) & = \frac{ \sqrt{N} }{ \rho_k \sqrt{ 2 \pi T ( 1 - c ) } }  \int_{-\infty}^{\hat{F}_k} d \hat{V}_k \ \left( 1 - \frac{ V_{k0} }{ F_k } \e{ \sqrt{z} \hat{V}_k +  \left( \mu_k - \frac{ \rho_k^2 }{ 2 } \right)T } \right)^j \notag \\ 
& \qquad \times \e{- \frac{ \left( \hat{V}_k + \sqrt{c T} u \rho_k \right)^2 }{ 2 T ( 1 - c) \rho_k^2 / N }  }
\label{eq:mjk}
\end{align}
and are given in appendix~\ref{ch:moments} for the case of a homogeneous portfolio with $j=0,1,2$. We carried out a change of variables to $\hat{V}_k=(\ln(V_k(T)/V_{k0}) - (\mu_k-\rho_k^2/2)T )/\sqrt{z}$ with the new upper bound for the integral
\begin{align}
	\hat{F}_k & = \frac{ 1 }{ \sqrt{z} } \left(\ln \frac{ F_k }{ V_{k0} } - \left( \mu_k - \frac{ \rho_k^2 }{ 2 } \right) T \right)  \quad 
\end{align}
The $z$ and $u$ integral must be evaluated numerically due to their complexity.

\subsection{Homogeneous portfolio}
\label{ch:homogen}

In case of a homogeneous portfolio all contracts of the portfolio have the same face value $F_k=F$, variance $\sigma^2_k=\sigma^2$, drift $\mu_k=\mu$ and start value $V_{k0}=V_0$. This makes the moment functions $m_{jk}(z,u)$ much faster to calculate as the $k$ dependence is dropped
\begin{align}
m_{j}(z,u) & = \frac{ \sqrt{N} }{ \rho \sqrt{ 2 \pi T ( 1 - c ) } }  \int_{-\infty}^{\hat{F}} d \hat{V} \left(  1 - \frac{V_{0} }{ F } \e{ \sqrt{z} \hat{V} +  \left( \mu - \frac{ \rho^2 }{ 2 } \right)T } \right)^j \notag \\
& \qquad \times \e{- \frac{ N \left( \hat{V} + \sqrt{cT} u \rho \right)^2 }{ 2 T ( 1 - c) \rho^2  }  }
\end{align}
with the upper limit for the integral
\begin{align}
\hat{F} & = \frac{ 1 } { \sqrt{z} } \left(\ln \frac{ F }{ V_{0} } - \left(\mu - \frac{ \rho^2 }{ 2 } \right) T \right) \quaddot
\end{align}
and $\hat{V}=(\ln(V(T)/V_{0}) - (\mu-\rho^2/2)T )/\sqrt{z}$.
Due to the normalization of the portfolio weights~(\ref{eq:def}) we have equal weights
\begin{align}
f_k = \frac{ 1 }{ K }
\label{eq:fraction}
\end{align}
for all contracts in the portfolio. The first three moments, which are required for $M_1$ and $M_2$, are shown in appendix~\ref{ch:moments}.

\subsection{Limit for very large homogeneous portfolios}
We calculate the limit case $K\rightarrow \infty$ to find out to what extent diversification can reduce the risk in a credit portfolio. The average loss distribution for the homogeneous case is
\begin{align}
\langle p \rangle (L|c,N) & = \frac{ 1 }{ \sqrt{ 2 \pi } 2^{N/2} \Gamma( N/2 ) } \int_0^\infty dz \ z^{N/2 -1} e^{-z/2}   \sqrt{ \frac{ N }{ 2 \pi } }  \int_{-\infty}^{+\infty} du \e{-\frac{ N }{ 2 } u^2 } \notag \\
& \qquad \times \frac{ 1 }{ \sqrt{ M_2(z,u) } } \e{ - \frac{ ( L - M_1(z,u) )^2 }{ 2 M_2(z,u) } } \quaddot
\end{align}
We notice that
\begin{align}
\frac{ 1 }{ \sqrt{ 2 \pi M_2(z,u) } } \e{ - \frac{ ( L - M_1(z,u) )^2 }{ 2 M_2(z,u) } }
\end{align}
becomes a delta function $\delta (L - M_1(z,u) )$ for $M_2(z,u) \rightarrow 0$. This happens in the homogeneous case,
\begin{align}
M_2(z,u) = \frac{ 1 }{ K } \left( m_{2}(z,u) - m_{1}^2(z,u) \right)
\end{align}
for $K \rightarrow \infty$, according to Eq.~(\ref{eq:M2}) and (\ref{eq:fraction}). We arrive at
\begin{align}
\langle p\rangle (L|c,N) & =  \frac{ 1 }{ 2^{N/2} \Gamma ( N/2 ) } \int_0^\infty dz \, z^{ N / 2 - 1 } e^{-z/2} \sqrt{ \frac{ N }{ 2 \pi } } \int_{-\infty}^{+\infty} du \, \exp \left( - \frac{ N }{ 2 } u^2 \right) \delta (L - m_1(z,u) ) \quaddot
\label{eq:withdelta}
\end{align}
Here, we use the integral form of the generalized scaling property
\begin{align}
	\int_{-\infty}^{+\infty} \text{d}u \, h(u) \delta( f(u) ) = \frac{ h(u_0) }{ \left| \partial f(u_0) / \partial u_0 \right| } 
\end{align}
for the $\delta$-function, where $u_0$ is the real root of the function $f(u)$.
The delta function in Eq.~(\ref{eq:withdelta}) yields only a contribution to the $u$ integral if its argument
\begin{align}
f(z,u) & = L-m_1(z,u)
\end{align}
is zero. We introduce the inverse function $u_0(L,z)$ according to
\begin{align}
0 & = L - m_1(z,u_0)
\label{eq:fzu}
\end{align}
and drop the arguments of $u_0$ to simplify the notation and determine the partial derivative
\begin{align}
\left| \frac{ \partial f(z,u) }{ \partial u } \right|_{z,u_0} & = \left| \frac{ \partial m_1 (z,u) }{ \partial u } \right|_{z,u_0}
\end{align}
at $u_0$.  This allows us to solve the $u$ integral using the generalized scaling property of the delta function. We arrive at the average loss distribution for the limit case $K \rightarrow \infty$
\begin{align}
\langle p \rangle(L|c,N) & =  \frac{ 1 }{ 2^{N/2} \Gamma( N/2) } \sqrt{ \frac{ N }{ 2 \pi } } \int_0^\infty dz \, z^{N/2-1} e^{-z/2} \exp \left( - \frac{ N }{ 2 } u_0^2 \right) \frac{ 1 }{  \left| \partial m_1 (z,u) / \partial u \right|_{z,u_0} }
\end{align}
We notice that the $L$ dependence is now in $u_0$ which is, according to Eq.~(\ref{eq:fzu}) a function of $L$ and $z$. The implications for credit portfolios will be discussed in section~\ref{ch:results}.

\section{Limits of stationary asset correlations}
\label{ch:results}

We present the average loss distribution in Sec.~\ref{ch:results:avg} for various combinations of empirically obtained parameters. In Sec.~\ref{ch:results:var} we study the Value at Risk and Expected Tail Loss of the average loss distribution.

\subsection{Average loss distribution}
\label{ch:results:avg}

We have shown in Sec.~\ref{ch:data} that the correlation averaged multivariate normal distribution~(\ref{eq:genresult}) is capable of describing the empirically observed return distribution. From this distribution we get the distribution of asset values~(\ref{eq:avg_result}) for the Merton model. The resulting distribution describes the asset values at maturity time $T$ and is entered into the Merton model~(\ref{eq:loss}).

Our model has four important parameters. Three of the parameters can be directly calculated from the empirical data, the average drift $\mu$, the average volatility $\sigma$ and the average correlation level $c$. The fourth parameter $N$ is proportional to the inverse variance of the elements of the Wishart correlation matrix, see~\cite{Chetalova2013} for a detailed description.
Therefore, $N$ controls the strength of the fluctuations around the average correlation level $c$ of the random matrix ensemble. We discussed the estimation of $N$ in Sec.~\ref{ch:data}.
\begin{figure*}[htbp]
  \begin{center}
    \includegraphics[width=0.95\textwidth]{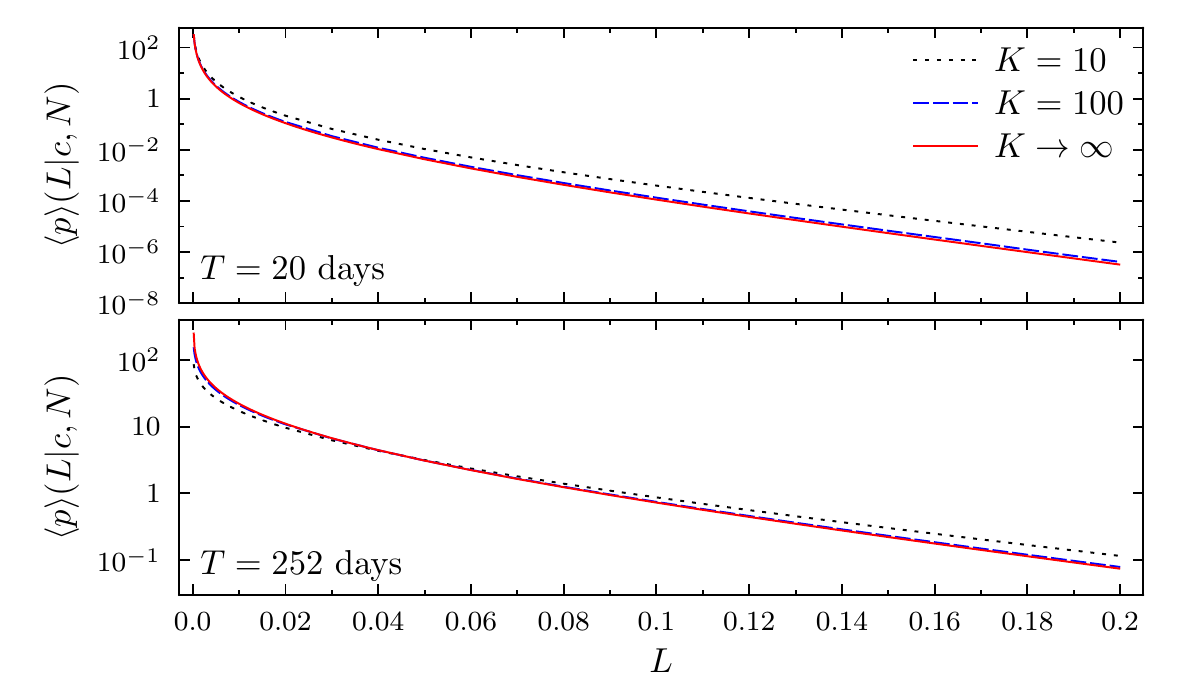}
  \end{center}
 \caption{The average loss distribution for portfolio sizes of $K=10$ and $K=100$. In addition the limit $K\rightarrow \infty$ is shown. The parameters are $N=4.2$, $\mu = 0.013$ month$^{-1}$, $\sigma=0.1$ month$^{-1/2}$, $T=20$ trading days and an average correlation level of $c=0.26$ (top); and $N=6.0$, $\mu = 0.17$ year$^{-1}$, $\sigma=0.35$ year$^{-1/2}$, a maturity time of $T=1$ year and an average correlation level of $c=0.28$ (bottom).}
 \label{fig:kdep}
\end{figure*}
Figure \ref{fig:kdep} shows the average loss distribution $\langle p \rangle(L|c,N)$ for correlation averaged asset values in the Merton model. We present the average loss distribution for typical empirical values of our parameters for different portfolio sizes $K=10,100$ and the limit $K\rightarrow \infty$. The face value is $F_0=75$ and the initial asset value at time $t=0$ is $V_0=100$. We notice the slowly decreasing heavy-tailed nature of the distribution. It can be clearly seen that increasing the size of the credit portfolio does not yield a significant decrease of the risk of large losses. Enlarging the portfolio from $10$ to $100$ contracts achieves a small decrease in risk. However, the distribution quickly converges to the limit $K\rightarrow \infty$, diminishing the effects of diversification. This contributes a quantitative reasoning why diversification is not working for credit portfolios in the presence of correlated assets. Starting from the empirical distribution of the multivariate returns we receive a data calibrated average loss distribution. 

\begin{figure*}[htbp]
  \begin{center}
    \includegraphics[width=0.95\textwidth]{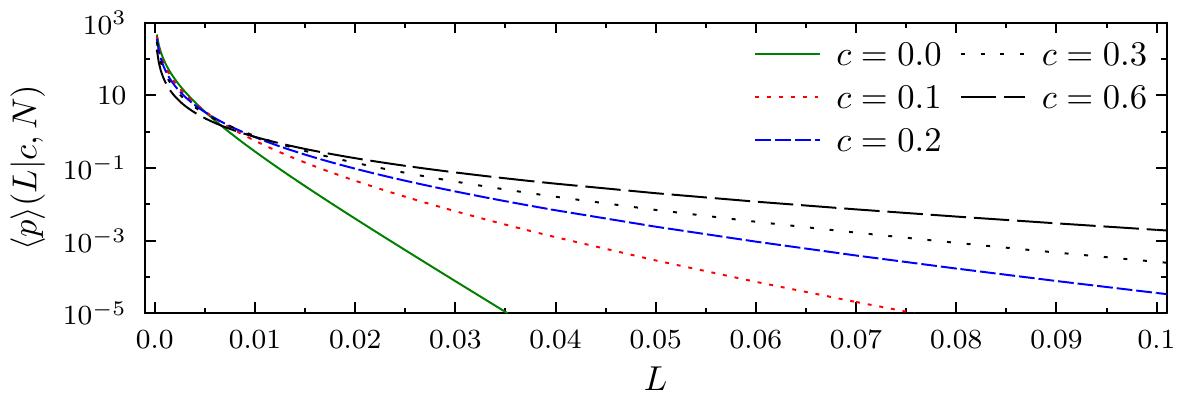}
  \end{center}
 \caption{Average loss distribution for different average correlation levels $c$. The parameters are $\mu = 0.013$ month$^{-1}$, $\sigma=0.1$ month$^{-1/2}$, $T=1$ month, $K=100$ and $N=4.2$.}
 \label{fig:cdep}
\end{figure*}
\begin{figure*}[htbp]
  \begin{center}
    \includegraphics[width=0.95\textwidth]{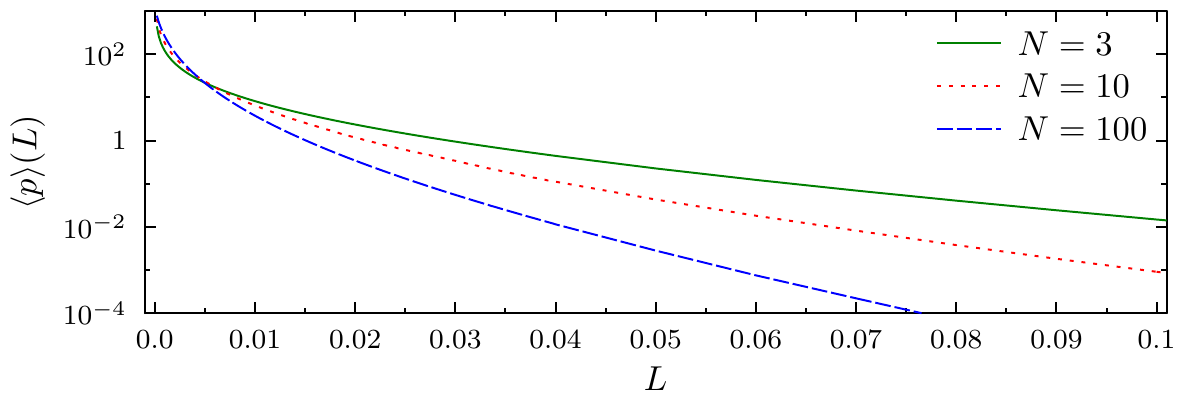}
  \end{center}
 \caption{Average loss distribution for different values of $N$. The parameters are $\mu = 0.015$ month$^{-1}$, $\sigma=0.25$ month$^{-1/2}$, $T=1$ month, $K=500$ and $c=0.2$.}
 \label{fig:ndep}
\end{figure*}

The effect of different average correlation levels $c$ is shown in Fig.~\ref{fig:cdep}. We observe that an increasing average correlation level leads to wider tails of the distribution, resulting in a higher risk of large losses.
The parameter $N$, which controls the strength of the fluctuations around the average correlation, plays an important role in calibrating the average loss distribution to the empirical returns. The dependence of the average loss distribution on $N$ is shown in Fig.~\ref{fig:ndep}. Smaller values of $N$ lead to a greater probability of large portfolio losses.
We emphasize that smaller values of $N$ correctly describe the data for our homogeneous correlation matrix with all off-diagonal elements equal to $c$.

\subsection{Value at Risk and Expected Tail Loss}
\label{ch:results:var}

In section \ref{ch:pavg}, we discussed the use of the homogeneous correlation matrix which has a fixed value $c$ on its off-diagonal elements. The homogeneous correlation matrix is necessary to make analytical progress. To study cases that extend the scope of the homogeneous correlation matrix, eg, empirical heterogeneous covariances, we use Monte-Carlo simulations to calculate the VaR and ETL. 
In each realization we calculate the $K$ dimensional vector 
\begin{align}
V(T) = V_0 \exp \left(  \frac{ \sqrt{T} }{ \sqrt{N} } \sigma  \left( U^{-1} \, \Lambda \, \mathcal{N} \right) n+ \mu T - \frac{ 1 }{ 2 }\sigma^2 e T  \right)
\end{align}
which contains the value of each asset at the maturity time $T$. $\mathcal{N}$ ist a $K\times N$ matrix whose elements are drawn from a standard normal distribution. The elements of the $N$ dimensional vector $n$ are drawn from the same distribution. The matrix $\sigma=\diag(\sigma_1,\dots, \sigma_K)$ contains the volatilities for each stock. The $K$ dimensional vector $\mu$ holds the drift for each stock and $e$ is a $K$ dimensional vector comprised of ones. The matrix $U$ contains the eigenvectors of the correlation matrix $C$ and the matrix $\Lambda$ holds the eigenvalues of the correlation matrix as diagonal elements. The $K$ dimensional vector $V_0$ contains the initial asset value at the beginning of the credit contract. Then the result of the simulation step is given by
\begin{align}
L^{(MC)} = \frac{ 1 }{ K } \sum_{k=0}^{K} \frac{ F_k - V_k(T) }{ F_k } \Theta ( F_k - V_k(T) ) \quadcomma
\end{align}
which is the portfolio loss. 
The theta function takes only cases into account where the asset value at maturity $V_k(T)$ is smaller than the face value $F_k$. We estimate the loss distribution as a histogram of the results for 50 million realizations.

First, we want to show that the homogeneous correlation matrix is quite capable in terms of estimating the VaR and the ETL. In financial applications it is not uncommon to estimate the covariance matrix over a longer period of time, say five years for example, and use it as an input for some risk estimation method. In this spirit, we want to know how well the homogeneous correlation matrix estimates the risk, taking fluctuating correlations into account. 
For different time horizons, we estimate the empirical covariance matrix for the monthly returns of the  S\&P 500 stocks. We compare the results for the empirical covariance matrix with the results for a covariance matrix with homogeneous correlation struture. For each time horizon we determine the parameter $N$ as described in section~\ref{ch:data}. In addition, we estimate the volatilities and drift for each stock and the average correlation $c$. The parameters are shown in Table~\ref{tab:params}. For volatility and drift we only show the average values $\bar{\sigma}$ and $\bar{\mu}$ over all stocks. Notice that $N$ must be an integer in our simulation. During the financial crisis a smaller value of $N_\text{emp}=7$ is necessary to model the higher than usual fluctuations of the volatilities.
\begin{table*}[htbp]
\centering
\begin{tabular}{rrrrrrrr}
\hline
Time horizon & $K$ & $N_\text{hom}$ & $N_\text{emp}$ & $\bar \sigma$ in & $\bar \mu$ in & $c$ \\ 
for estimation &  &  &  & month$^{-1/2}$ & month$^{-1}$  &  & \\ 
\hline
2006-2010 & 465 & 5 & 12 & 0.11 & 0.009 & 0.40\\
2002-2004 & 436 & 5 & 14 & 0.10 & 0.015 & 0.30\\
2008-2010 & 478 & 5 & 7 & 0.12 & 0.01 & 0.46 \\ 
\hline
\end{tabular}
\caption{Parameters used for the different time horizons.}
\label{tab:params}
\end{table*}

We calculate the relative deviation of the VaR and ETL for different quantiles $\alpha=0.99, 0.995, 0.999$ from the empirical covariance matrix. We study two cases for the covariance matrix with homogeneous correlation structure. First, we use the average values of volatility and drift for each stock. This resembles the homogeneous case discussed in section~\ref{ch:homogen} and is shown in Table~\ref{tab:var_mod}. Second, we use the empirically obtained volatilities and drifts for each stock, see Table~\ref{tab:var_emp}. Positive values of the relative deviation indicate that the covariance matrix with homogeneous correlation structure overestimates VaR and ETL, while negative values show an underestimation. We round all values to an accuracy of $0.5$. For homogeneous volatilities and drifts we find that the covariance matrix with homogeneous correlations underestimates the risk in most cases. If we use heterogeneous volatilities and drifts, we find that the covariance matrix with homogeneous correlations is an appropriate fit and in most cases slightly overestimates the VaR and ETL. 
\begin{table*}[htbp]
\centering
\begin{tabular}{crrrrrrr}
\hline
time& $F/V_0$ & $\delta_{\text{VaR},{99.0}}$ & $\delta_{\text{VaR},{99.5}}$ & $\delta_{\text{VaR},{99.9}}$ & $\delta_{\text{ETL},{99.0}}$ & $\delta_{\text{ETL},{99.5}}$ & $\delta_{\text{ETL},{99.9}}$\\ 
horizon &  & in \%  & in \%  & in \%  & in \%  & in \%  & in \% \\ 
\hline
2006-2010 & 0.75 & -45.5 & -35.0 & -13.5 & -26.0 & -18.0 & -1.0\\
 & 0.80 & -21.0 & -12.5 & 3.0 & -6.5 & -0.5 & 11.5\\
 & 0.85 & -4.5 & 1.0 & 11.0 & 4.0 & 8.0 & 15.5\\
 & 0.90 & 3.0 & 6.5 & 12.0 & 8.0 & 10.5 & 15.0\\ 
2002-2004 & 0.75 & -69.5 & -60.0 & -38.0 & -51.0 & -42.5 & -22.0\\ 
 & 0.80 & -41.0 & -30.5 & -9.0 & -23.0 & -14.0 & 3.5\\ 
 & 0.85 & -12.5 & -4.0 & 11.0 & 0.5 & 7.0 & 18.5\\ 
 & 0.90 & 4.5 & 9.0 & 17.5 & 11.5 & 15.0 & 22.0\\ 
2008-2010 & 0.75 & -42.5 & -33.5 & -17.0 & -27.0 & -21.0 & -8.5\\ 
 & 0.80 & -21.5 & -15.5 & -4.0 & -11.5 & -7.0 & 1.0\\ 
 & 0.85 & -8.0 & -4.0 & 2.5 & -2.0 & 0.5 & 5.0\\ 
 & 0.90 & -1.0 & 1.0 & 5.0 & 2.0 & 3.5 & 6.5\\ 
\hline
\end{tabular}
\caption{Relative deviation $\delta$ of the VaR and ETL between the homogeneous correlation matrix and an empirical covariance matrix in percent. We use homogeneous volatility and drift vectors. Positive values denote that the covariance matrix with homogeneous correlations overestimates VaR and ETL, while negative values show an underestimation. We present the VaR and ETL at 99\%, 99.5\% and 99.9\%.}
\label{tab:var_mod}
\end{table*}
 \begin{table*}[htbp]
\centering
\begin{tabular}{crrrrrrr}
\hline
time& $F/V_0$ & $\delta_{\text{VaR},{99.0}}$ & $\delta_{\text{VaR},{99.5}}$ & $\delta_{\text{VaR},{99.9}}$ & $\delta_{\text{ETL},{99.0}}$ & $\delta_{\text{ETL},{99.5}}$ & $\delta_{\text{ETL},{99.9}}$\\ 
horizon &  & in \%  & in \%  & in \%  & in \%  & in \%  & in \% \\ 
\hline
2006-2010 & 0.75 & 18.0 & 18.5 & 20.0 & 19.5 & 20.0 & 22.5\\
 & 0.80 & 12.0 & 13.0 & 16.0 & 14.0 & 15.0 & 18.0\\
 & 0.85 & 7.5 & 9.0 & 12.5 & 10.0 & 12.0 & 15.0\\
 & 0.90 & 5.0 & 6.5 & 10.5 & 8.0 & 9.5 & 12.5\\
2002-2004 & 0.75 & 12.0 & 14.0 & 19.5 & 16.5 & 18.5 & 24.0\\ 
 & 0.80 & 12.0 & 14.5 & 20.5 & 17.0 & 19.0 & 24.5\\ 
 & 0.85 & 11.5 & 14.5 & 20.0 & 16.0 & 18.5 & 23.0\\
 & 0.90 & 10.0 & 12.5 & 17.0 & 14.0 & 15.5 & 19.5\\ 
2008-2010 & 0.75 & -1.0 & -1.5 & -0.50 & -1.0 & -1.0 & 0.5\\ 
 & 0.80 & -2.0 & -2.0 & -0.0 & -1.0 & -0.5 & 1.5\\ 
 & 0.85 & -2.0 & -1.5 & 1.0 & -0.5 & 0.0 & 2.0\\ 
 & 0.90 & -1.5 & -0.5 & 1.5 & 0.0 & 1.0 & 3.0\\ 
\hline
\end{tabular}
\caption{Relative deviation $\delta$ of the VaR and ETL between the homogeneous correlation matrix and an empirical covariance matrix in percent. We use the empirical volatilities and drifts for each stock. Positive values denote that the covariance matrix with homogeneous correlations overestimates VaR and ETL, while negative values show an underestimation. We present the VaR and ETL at 99\%, 99.5\% and 99.9\%.}
\label{tab:var_emp}
\end{table*}
In all cases we observe decreasing deviations from the empirical covariance matrix for larger leverages $F/V_0$. 
This shows that the structure of the correlation matrix plays a minor role and underlines the importance of getting the volatilities right.

\begin{figure*}[htbp]
  \begin{center}
    \includegraphics[width=0.95\textwidth]{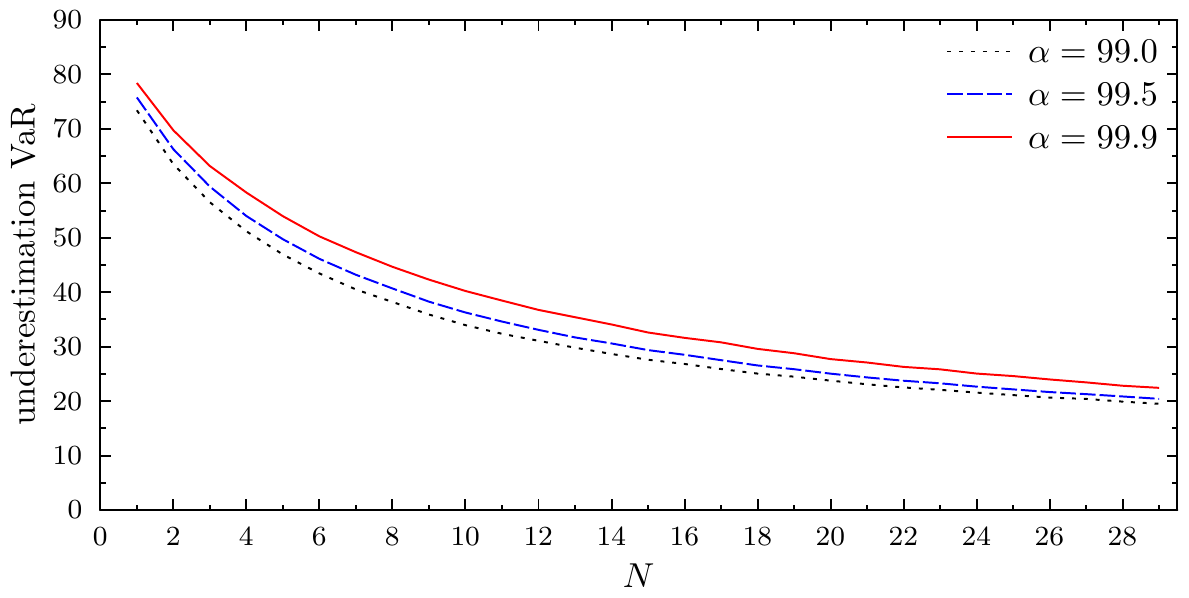}
  \end{center}
 \caption{Underestimation of the VaR for the empirical covariance matrix (2006-2010) if fluctuating correlations are not taken into account. Comparison for different values of $N$. The empirical  value for this case is $N_\text{emp}=12$.}
 \label{fig:var}
\end{figure*}

In Figure~\ref{fig:var} we demonstrate how the VaR is underestimated by using stationary correlations. We calculate the relative deviation of the VaR for $N \rightarrow \infty$ and for different values of $N$. The case $N \rightarrow \infty$ does not allow fluctuations of the covariance matrix and thereby effectively disables the benefits of the random matrix approach. We use the empirical covariance matrix for 2006-2010, ie, the volatilities and drifts are heterogeneous. For the empirically observed value of $N=12$ the VaR is underestimated between $30$\% and $40$\%. We recall that an empirical covariance matrix needs larger values for the parameter $N$ as discussed in section~\ref{ch:data}.

\begin{figure*}[htbp]
  \begin{center}
    \includegraphics[width=0.95\textwidth]{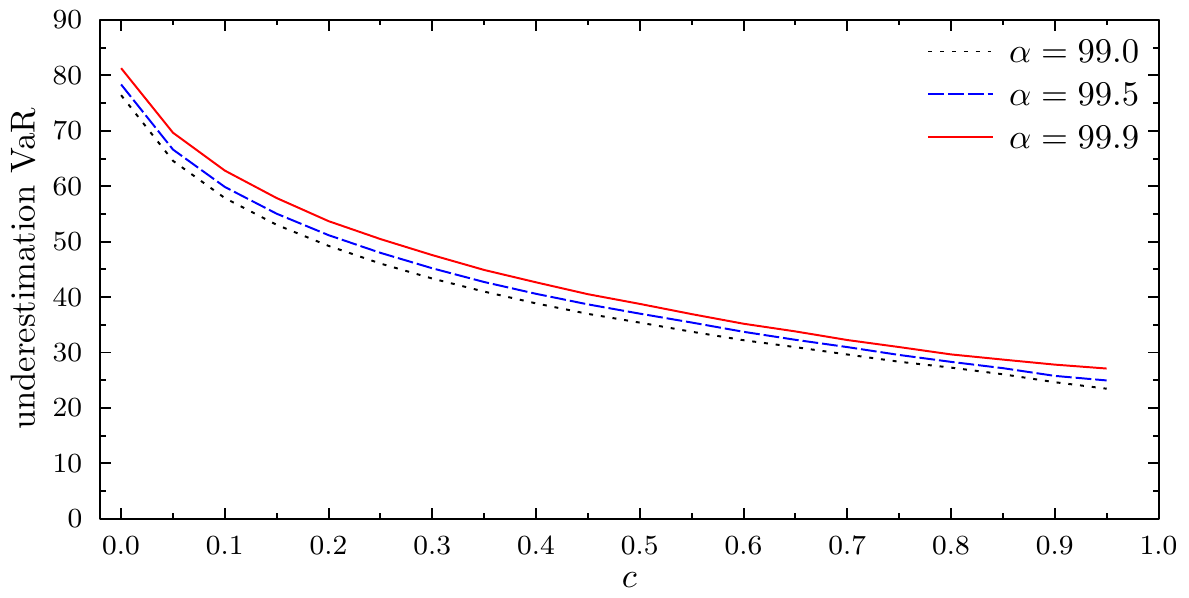}
  \end{center}
 \caption{Underestimation of the VaR in case of the homogeneous correlation matrix if fluctuating correlations are not taken into account. We use homogeneous volatilities and drifts. Parameters are $N=5$, $K=500$, $\bar \sigma=0.25$ year$^{-1/2}$, $\bar \mu = 0.15$ year$^{-1}$ and different values of $c$. }
 \label{fig:mod}
\end{figure*}

For a homogeneous correlation matrix, Figure~\ref{fig:mod} shows how much stationary correlations underestimate the VaR dependent on the average correlation level $c$. The VaR is underestimated by roughly 45\% for typical average correlation levels between $0.2$ and $0.4$.

\section{Conclusion}

We showed that an ensemble average of random correlation matrices yields a suitable asset value distribution. While empirical data on asset values is difficult to obtain, we can use stock prices as a good proxy according to the Merton model.
Our ensemble approach is supported by two different comparisons with empirical data. 
First, we showed that stock returns follow a multivariate normal distribution, if the time interval for the sample is rather short. Second, the sample statistics of returns on large time horizons can be described by a multivariate mixture, where the multivariate normal distribution is averaged over an ensemble of Wishart-distributed covariance matrices.

This allowed us to derive the loss distribution for a portfolio of credit contracts, taking fluctuating correlations between asset values into account. In addition, we were able to derive an analytical expression for the limit distribution of infinite portfolio size in case of a homogeneous portfolio. 

We performed Monte-Carlo simulations for VaR and ETL. The results support our ansatz of homogeneous average correlations if the heterogeneous average volatilities are taken into account. The simulations  reveal an underestimation of the VaR by roughly 40\% when fluctuating correlations between the asset values are neglected. 
Our ensemble approach allows us to capture the correlation structure of the financial market by only two ``macroscopic'' parameters, the average correlation level and the strength of the fluctuations around this average.
In addition, our model provides a quantitative understanding for why the presence of asset correlations severely limits the benefits of diversification for credit risk. 
As a consequence, our results strongly support a conservative approach to capital reserve requirements.

\section{Acknowledgments}
We thank Desislava Chetalova and Michael C. M\"unnix for fruitful discussions.

\bibliographystyle{unsrt}

\begin{thebibliography}{10}

\bibitem{Hull2009}
John~C. Hull.
\newblock {The credit crunch of 2007: What went wrong? Why? What lessons can be
  learned?}
\newblock {\em The Journal of Credit Risk}, 5(2):3--18, 2009.

\bibitem{Crouhy2000}
Michel Crouhy, Dan Galai, and Robert Mark.
\newblock {A comparative analysis of current credit risk models}.
\newblock {\em Journal of Banking \& Finance}, 24(1-2):59--117, 2000.

\bibitem{bielecki2004credit}
T.~R. Bielecki and M.~Rutkowski.
\newblock {\em {Credit Risk: Modeling, Valuation and Hedging}}.
\newblock Springer, 2004.

\bibitem{bluhm2003introduction}
C.~Bluhm, L.~Overbeck, and C.~Wagner.
\newblock {\em {An Introduction to Credit Risk Modeling}}.
\newblock Chapman \& Hall/CRC, 2003.

\bibitem{Duffie1999}
Darrell Duffie and Kenneth~J. Singleton.
\newblock {Modeling Term Structures of Defaultable Bonds}.
\newblock {\em Review of Financial Studies}, 12(4):687--720, 1999.

\bibitem{Ibragimov2007}
Rustam Ibragimov and Johan Walden.
\newblock {The limits of diversification when losses may be large}.
\newblock {\em Journal of Banking \& Finance}, 31(8):2551--2569, 2007.

\bibitem{lando2008credit}
D.~Lando.
\newblock {\em {Credit Risk Modeling: Theory and Applications}}.
\newblock Princeton University Press, 2008.

\bibitem{mcneil2005quantitative}
A.~J. McNeil, R.~Frey, and P.~Embrechts.
\newblock {\em {Quantitative Risk Management: Concepts, Techniques, and
  Tools}}.
\newblock Princeton University Press, 2005.

\bibitem{Heitfield2006}
Erik Heitfield, Steve Burton, and Souphala Chomsisengphet.
\newblock {Systematic and idiosyncratic risk in syndicated loan portfolios}.
\newblock {\em Journal of Credit Risk}, 2(3):3--31, 2006.

\bibitem{Glasserman2006}
Paul Glasserman and Jesus Ruiz-Mata.
\newblock {Computing the credit loss distribution in the Gaussian copula model:
  a comparison of methods}.
\newblock {\em Journal of Credit Risk}, 2(4):33--66, 2006.

\bibitem{Mainik2012}
Georg Mainik and Paul Embrechts.
\newblock {Diversification in heavy-tailed portfolios: properties and
  pitfalls}.
\newblock {\em Annals of Actuarial Science}, 7(1):26--45, 2013.

\bibitem{Schonbucher2001}
Philipp~J. Sch\"{o}nbucher.
\newblock {Factor Models: Portfolio Credit Risks When Defaults are Correlated}.
\newblock {\em The Journal of Risk Finance}, 3(1):45--66, 2001.

\bibitem{Glasserman2004}
Paul Glasserman.
\newblock {Tail approximations for portfolio credit risk}.
\newblock {\em The Journal of Derivatives}, 12(2):24--42, 2004.

\bibitem{Schafer2007}
Rudi Sch\"{a}fer, Markus Sj\"{o}lin, Andreas Sundin, Michal Wolanski, and
  Thomas Guhr.
\newblock {Credit risk - A structural model with jumps and correlations}.
\newblock {\em Physica A}, 383(2):533--569, September 2007.

\bibitem{Koivusalo2011}
Alexander F.~R. Koivusalo and Rudi Sch\"{a}fer.
\newblock {Calibration of structural and reduced-form recovery models}.
\newblock {\em Journal of Credit Risk}, 8(4):31--51, 2012.

\bibitem{Schmitt2014}
Thilo~A. Schmitt, Desislava Chetalova, Rudi Sch\"{a}fer, and Thomas Guhr.
\newblock {Credit risk and the instability of the financial system: An ensemble
  approach}.
\newblock {\em Europhysics Letters}, 105:38004, 2014.

\bibitem{Schmitt2013}
Thilo~A. Schmitt, Desislava Chetalova, Rudi Sch\"{a}fer, and Thomas Guhr.
\newblock {Non-Stationarity in Financial Time Series and Generic Features}.
\newblock {\em Europhysics Letters}, 103:58003, 2013.

\bibitem{Zhang2011}
Yiting Zhang, Gladys Hui~Ting Lee, Jian~Cheng Wong, Jun~Liang Kok, Manamohan
  Prusty, and Siew~Ann Cheong.
\newblock {Will the US economy recover in 2010? A minimal spanning tree study}.
\newblock {\em Physica A}, 390(11):2020--2050, June 2011.

\bibitem{Song2011}
Dong-Ming Song, Michele Tumminello, Wei-Xing Zhou, and Rosario~N. Mantegna.
\newblock {Evolution of worldwide stock markets, correlation structure, and
  correlation-based graphs}.
\newblock {\em Physical Review E}, 84(2):026108, August 2011.

\bibitem{Munnix2012}
Michael~C. M\"{u}nnix, Takashi Shimada, Rudi Sch\"{a}fer, Francois Leyvraz,
  Thomas~H Seligman, Thomas Guhr, and H.~Eugene Stanley.
\newblock {Identifying states of a financial market.}
\newblock {\em Scientific reports}, 2:644, January 2012.

\bibitem{Sandoval2012}
Leonidas Sandoval and Italo De~Paula Franca.
\newblock {Correlation of financial markets in times of crisis}.
\newblock {\em Physica A}, 391(1-2):187--208, January 2012.

\bibitem{Chetalova2013}
Desislava Chetalova, Thilo~A. Schmitt, Rudi Sch\"{a}fer, and Thomas Guhr.
\newblock {Portfolio return distributions: Sample statistics with
  non-stationary correlations}.
\newblock {\em ArXiv: 1308.3961}, 2013.

\bibitem{Munnix2011a}
Michael~C. M\"{u}nnix, Rudi Sch\"{a}fer, and Thomas Guhr.
\newblock {A Random Matrix Approach on Credit Risk}.
\newblock {\em PLOS ONE}, 9(5), 2014.

\bibitem{Merton1974}
Robert~C. Merton.
\newblock {On the pricing of corporate debt: the risk structure of interest
  rates}.
\newblock {\em The Journal of Finance}, 29(2):449--470, 1974.

\bibitem{Wishart1928}
J~Wishart.
\newblock {The Generalised Product Moment Distribution in Samples from a Normal
  Multivariate Population}.
\newblock {\em Biometrika}, 20A(1/2):32--52, 1928.

\bibitem{Ito1944}
Kiyosi It\^{o}.
\newblock {Stochastic integral}.
\newblock {\em Proceedings of the Imperial Academy}, 20(8):519--524, 1944.

\bibitem{yahoo}
{Standard \& Poor's 500 data from Yahoo! Finance}, 2013.

\end{thebibliography}

\clearpage 
\appendix

\section{Alternative estimation of the parameter N}
\label{ch:appendix:var}

We discuss an additional approach to estimate the parameter $N$ for homogeneous average correlations. According to our model the returns
\begin{align}
r = W \varepsilon
\end{align}
can be written as a product of the random $K \times N$ matrix $W$ with independent rows and a stochastic vector $\varepsilon$. The $N$ dimensional vector $\varepsilon$ consists of \textit{i.i.d.} normal distributed random numbers with zero mean and variance one. We calculate the variance of the expression
\begin{align}
x = \trace r r^\dagger = r^\dagger r = \sum_{kij} W_{ki} \varepsilon_i W_{kj} \varepsilon_j \quadcomma
\end{align}
which yields
\begin{align}
\langle x^2 \rangle = 4 \left(\frac{ 1 }{ 2 } + c^2 \right) \frac{ K^2 }{ N } + 2 c^2 K^2 + 4(1-c^2)  \frac{ K }{ N }+2(1-c^2) K 
\end{align}
The variance can be directly compared to the variance of empirical return data. For both of our data sets on the above time horizons we find that values of $N$ smaller than $5$ are necessary to reproduce the empirical variance.

\section{Moments}
\label{ch:moments}
We use the following definition
\begin{align}
\Phi(x)  & = \frac{ 1 }{ 2 } + \frac{ 1 }{ 2 } \erf \left( \frac{ x }{ \sqrt{2} } \right)
\end{align}
with the error function
\begin{align}
	\erf(y) = \frac{ 2 }{ \sqrt{\pi} } \int_0^y \text{d} \tau e^{-\tau^2}
\end{align}
to express the moments.
\subsubsection*{Zeroth moment}
\begin{align}
m_0(u) = & \int_{-\infty}^{ \hat{F} } d\hat{V} \sqrt{ \frac{ N }{ 2 \pi T (1-c) \rho^2 } } \e{ - \frac{ (\hat{V} + \sqrt{cT} u \rho )^2 }{ 2 T (1-c) \rho^2 / N } } \notag \\
 = & \Phi \left( \frac{ \hat{F} + \sqrt{cT} u \rho }{ \sqrt{ T (1-c) \rho^2 /N } } \right)
 \end{align}
 \subsubsection*{First moment}
 \begin{align}
m_1(z,u) = & m_0(u)- \frac{ V_0 }{ F } \int_{-\infty}^{\hat{F}} d\hat{V} \sqrt{ \frac{ N }{ 2 \pi T (1-c) \rho^2 } } \notag \\
  & \qquad \times \e{ \sqrt{z} \hat{V} + (\mu -\frac{ \rho^2 }{ 2 } ) T} \e{ - \frac{ (\hat{V} + \sqrt{cT} u \rho )^2 }{ 2T (1-c) \rho^2 / N } } \notag \\
  = & m_0(z,u)- \frac{ V_0 }{ F } \e{ (\mu -\frac{ \rho^2 }{ 2 } ) T -  \sqrt{cTz} u \rho + T(1-c) \frac{ \rho^2 z }{ 2 N }  } \notag \\
& \qquad\times \Phi \left( \frac{ \hat{F} - T (1-c)\rho^2 \sqrt{z}/N + \sqrt{c T} u \rho }{ \sqrt{ T (1-c) \rho^2 / N } } \right)
\end{align}
 \subsubsection*{Second moment}
\begin{align}
m_2(z,u) & = - m_0(u) + 2 m_1(z,u)  + \frac{ V_0^2 }{ F^2 } \int_{-\infty}^{\hat{F}} d\hat{V} \sqrt{ \frac{ N }{ 2 \pi T (1-c) \rho^2 } } \notag \\
& \qquad \times \e{ 2 \sqrt{z} \hat{V} + 2 \mu T -  \rho^2 T}  \e{ - \frac{ (\hat{V} + \sqrt{cT} u \rho )^2 }{ 2T (1-c) \rho^2 / N } } \notag \\
& = - m_0(z,u) + 2 m_1(z,u)  +   \frac{ V_0^2 }{ F^2 } \e{ 2 \mu T -  \rho^2 T - 2 \sqrt{c T z} u \rho - 2T(1-c) \frac{ \rho^2 z }{ N } } \notag \\
&\qquad \times \Phi \left( \frac{ \hat{F} + \sqrt{cT} u \rho - 2T (1-c) \rho^2 \sqrt{z} /N}{ \sqrt{ T (1-c) \rho^2 / N } } \right)
\end{align}

\end{document}